\documentclass[aps,prd,reprint,preprintnumbers,showpacs,showkeys,superscriptaddress]{revtex4-1}
\usepackage{graphicx}
\usepackage{amsmath}
\usepackage{bm}
\usepackage{slashed}

\newcommand{\Sqrt}[1]{\sqrt{\mathstrut #1}}
\newcommand\vek[1]{\bm{#1}} 
\newcommand\he[1]{#1^{\dagger}} 
\newcommand\gr[1]{\mathrm{#1}} 
\DeclareMathOperator{\tr}{tr}

\begin{document}

\title{How does color neutrality affect collective modes in color superconductors?}

\author{Hiroaki Abuki}
\email{h.abuki@rs.kagu.tus.ac.jp}
\affiliation{Department of Physics, Tokyo University of Science, Kagurazaka 1--3, Shinjuku, Tokyo 162--8601, Japan}

\author{Tom\'{a}\v{s} Brauner}
\email{tbrauner@physik.uni-bielefeld.de}
\affiliation{Faculty of Physics, University of Bielefeld, 33615 Bielefeld, Germany} 
\affiliation{Department of Theoretical Physics, Nuclear Physics Institute ASCR, 25068 \v Re\v z, Czech Republic}

\begin{abstract}
We revisit the issue of color neutrality in effective model descriptions of dense quark matter based on global color symmetry. While the equilibrium thermodynamics of such models is now well understood, we examine the collective modes, focusing on the fluctuations of the order parameter. We point out that the constraint of color neutrality must be carefully generalized in order to obtain physically consistent and well-defined results. Particularly important is that the collective modes associated with order parameter fluctuations couple to charge density fluctuations in the neutral medium. We start by proving explicitly that, in contrast to claims made previously in literature, Nambu--Goldstone bosons of spontaneously broken global color symmetry remain exactly massless even after imposing the color neutrality constraint. As the next step, we make the argument general by using effective field theory. We then employ the high-density approximation to calculate the couplings in the effective Lagrangian and thus the Nambu--Goldstone boson dispersion relations.
\end{abstract}

\preprint{BI-TP 2012/07}

\maketitle


\section{Introduction}

The physics of cold dense quark matter is governed by the theory of strong interactions, the Quantum Chromodynamics (QCD). Unfortunately, it becomes strongly coupled in the phenomenologically interesting range of densities. At present, there is no analytic method in the market that would be able to perform reliable first-principle calculations in this regime, although first attempts have already been made~\cite{Kurkela:2009gj}. Likewise, lattice simulations at high density are out of reach of the standard Monte-Carlo techniques due to the sign problem.

As a consequence, one usually has to resort to simplified calculations using models which more or less imitate the full QCD. Since the technical difficulty of QCD stems from the strong gauge interaction, these models invariably replace the color gauge symmetry with a global one, the prototype being the Nambu--Jona-Lasinio (NJL) model (see Ref.~\cite{Vogl:1991qt,*Klevansky:1992qe,*Hatsuda:1994pi,*Buballa:2003qv} for extensive reviews). One then has to deal with various model artifacts that are particularly severe at high baryon density where cold quark matter is expected to behave as a color superconductor (see Ref.~\cite{Alford:2007xm,*Wang:2009xf,*Huang:2010nn,*Fukushima:2010bq} for recent reviews). A condensate of quark Cooper pairs breaks color symmetry and (in most color-superconducting phases) induces nonzero color charge density. This cannot be physical since, due to the long-range nature of gauge interactions, it would give rise to nonextensive energy density and an ill-defined thermodynamic limit~\cite{Alford:2002kj}. In QCD, it is compensated by an induced gluon condensate so that the system as a whole is color neutral~\cite{Gerhold:2003js,*Dietrich:2003nu}. On the other hand, in models with global color symmetry color neutrality has to be imposed as a thermodynamic constraint.

Arranging for global color neutrality by introducing one or more chemical potentials associated with the color charge(s) costs energy as compared to the unconstrained equilibrium state. The neutrality constraint then has to be defined carefully in order to avoid spurious instabilities. Initially, only neutrality with respect to the three colors of fundamental quarks was required in literature on color superconductivity, leading to the existence of seemingly neutral and energetically preferred states~\cite{He:2005jq,*Blaschke:2005km}. As was pointed out in Ref.~\cite{Buballa:2005bv}, such a restricted neutrality requirement is sufficient only for special orientations of the diquark condensate in the color space. In general, neutrality with respect to all eight generators of color $\gr{SU(3)}$ has to be imposed, and the full set of eight color chemical potentials then have to be introduced~\footnote{This is in no contradiction with the common knowledge that only mutually commuting charges can be simultaneously fixed in a grand canonical ensemble, giving rise to just two independent chemical potentials for $\gr{SU(3)}$, as dictated by its rank.   In fact, one can always choose a basis in the Lie algebra of $\gr{SU(3)}$ such that only two color charges are nonzero. In such a basis, two chemical potentials are sufficient to make the system neutral. However, even in this case, the other chemical potentials will be important in our discussion of order parameter fluctuations.}.

Once the chemical potentials are fixed to make the equilibrium (``ground state'') neutral, color symmetry is apparently broken explicitly. This observation led to the conclusion that the Nambu--Goldstone (NG) bosons of the spontaneously broken global color symmetry have small, yet nonzero masses, proportional to the color chemical potential(s)~\cite{He:2005mp,*Ebert:2005fi,*Ebert:2006bq}. The goal of the present paper is to show that this conclusion is premature. The physical picture behind our claim is as follows. The introduction of chemical potentials is enforced by the diquark condensate, which itself is generated dynamically as a non-perturbative solution to the equations of motion. The color chemical potentials therefore are not mere external fields, but stem from the dynamics of the system. In full QCD, this is indeed the case~\cite{Gerhold:2003js,*Dietrich:2003nu}. After all, none of the color symmetry is broken explicitly once the dynamics in the gauge sector is properly taken into account. In the NJL model, the color chemical potentials mimic the role of the gluon condensate in the full QCD.

One should note that the chemical potentials commonly used in literature are demanded to make the \emph{ground state} neutral. However, the NG bosons constitute its excitations. One therefore cannot use the very same values of the chemical potentials when dealing with such non-equilibrium field configurations, and the thermodynamic constraint of color neutrality needs to be extended properly. We show that this can be technically achieved by treating the chemical potentials as dynamical variables, or in other words, secondary order parameters in the gauge sector induced by the primary order parameter, that is, the diquark condensate. Once this is done, the NG bosons remain exactly massless, as predicted by the Goldstone theorem.

The plan of the paper is as follows. In Section~\ref{Sec:toymodel} we explain the central idea of the paper using a very simple toy model. While this is rather trivial, it is intended to demonstrate the conceptual simplicity of our strategy, which might otherwise be concealed by unimportant technical details of the NJL calculation to follow. Section~\ref{Sec:NGmasses} provides a correction of the calculation of the NG boson masses of Ref.~\cite{He:2005mp,*Ebert:2005fi,*Ebert:2006bq}, showing that they are exactly zero once color neutrality is properly imposed. In Section~\ref{Sec:NGdispersions} we extend the calculation and determine the full dispersion relations of the NG bosons. We employ the high-density approximation~\cite{Evans:1998ek,*Hong:1998tn,*Hong:1999ru} (see Ref.~\cite{Nardulli:2002ma} for a review and further references), which both simplifies the calculation and makes the results model independent, for it is known to capture the leading order of the high-density asymptotic behavior in full QCD. Finally, in Section~\ref{Sec:conclusions} we summarize and conclude.


\section{Neutrality in a scalar toy model}
\label{Sec:toymodel}

Consider the scalar theory with a global $\gr{SU(2)\times U(1)}$ symmetry, defined by the (Minkowski space) Lagrangian
\begin{equation}
\mathcal L=\mathcal D_\mu\he\phi \mathcal D^{\mu}\phi-M^2\he\phi\phi-\lambda(\he\phi\phi)^2,
\label{Lagrangian}
\end{equation}
where $\phi$ is a complex doublet field. This model was investigated many times before~\cite{Miransky:2001tw,*Schaefer:2001bq,*Andersen:2005yk} and it was shown that when the symmetry is spontaneously broken in presence of a chemical potential associated with the $\gr{U(1)}$ subgroup, the three broken generators give rise to one type-I and one type-II NG boson whose dispersion relations at low momentum are linear and quadratic, respectively~\footnote{This conclusion does not depend on whether the symmetry is actually broken by the chemical potential itself, or rather by $M^2<0$ as in the Higgs model, and the chemical potential only introduces medium effects.}. This is in accordance with the general Nielsen--Chadha counting rule for the number of NG bosons as well as the fact that in the ground state, isospin acquires nonzero density~\cite{Nielsen:1975hm,*Watanabe:2011ec,*Watanabe:2012,*Hidaka:2012}.

Here we wish to investigate the effect of enforcing neutrality with respect to the isospin $\gr{SU(2)}$ group. This may be regarded as a simple toy model for understanding charge neutrality in non-Abelian gauge theories where the global symmetry (after gauge fixing) is spontaneously broken such as in color superconductors. We will explain that using the values of chemical potentials obtained from the neutrality constraint on the ground state to determine the excitation spectrum leads to a spurious instability. We will then use two different approaches to deal with it. First, we shall demand, and justify, that all uniform field configurations be $\gr{SU(2)}$ neutral, whether they correspond to the thermodynamic equilibrium or not. A more model-independent argument will be given afterwards, showing that the $\gr{SU(2)}$ chemical potentials can be treated as induced, secondary order parameters. These two approaches will be later used respectively in Sections~\ref{Sec:NGmasses} and~\ref{Sec:NGdispersions}. Here we just remark that while the former has the advantage of being conceptually more straightforward, the latter is more powerful and allows us in particular to determine the full dispersion relations of the NG bosons.


\subsection{Spurious instability and its cure}
\label{Subsec:instability}

We start by introducing independent chemical potentials for all generators of the symmetry group, $\mu$ for $\gr{U(1)}$ and $\vec\mu$ for $\gr{SU(2)}$. The covariant derivative in Eq.~\eqref{Lagrangian} then reads
\begin{equation}
\mathcal D_\mu\phi=[\partial_\mu-i\delta_{\mu0}(\mu+\vec\tau\cdot\vec\mu)]\phi,
\end{equation}
where $\vec\tau$ is the vector of Pauli matrices. The presence of chemical potentials for all charges allows us to easily find the associated charge density operators (here for the isospin charges),
\begin{equation}
\vec n=\frac{\partial\mathcal L}{\partial\vec\mu}=i\left(\he\phi\vec\tau\partial_0\phi-\partial_0\he\phi\vec\tau\phi\right)+2\he\phi(\mu\vec\tau+\vec\mu)\phi.
\label{charge_density}
\end{equation}
Using this expression, the Lagrangian \eqref{Lagrangian} becomes
\begin{equation}
\begin{split}
\mathcal L=&\partial_\mu\he\phi\partial^\mu\phi+i\mu(\he\phi\partial_0\phi-\partial_0\he\phi\phi)+\vec\mu\cdot\vec n\\
&-\he\phi(M^2-\mu^2+\vec\mu^2)\phi-\lambda(\he\phi\phi)^2.
\end{split}
\label{Lagrangian_expanded}
\end{equation}
All matrix structure is now hidden in the density $\vec n$.

Let us assume that the scalar field develops nonzero vacuum expectation value (the conditions for this to happen will be discussed below) and choose the ground state as usual as $\phi_0=(0,v)^T$. The complex doublet field will thus be parameterized as $\phi_1=\varphi$, $\phi_2=v+H+i\theta$, where $\varphi$ is a complex field, whereas $H$ and $\theta$ are real. For constant field configurations, the charge densities~\eqref{charge_density} are then expressed as
\begin{equation}
\begin{split}
\frac12n_1&=\mu_1(v^2+2vH)+2\mu v\,\mathrm{Re}\,\varphi+\text{second-order terms},\\
\frac12n_2&=\mu_2(v^2+2vH)-2\mu v\,\mathrm{Im}\,\varphi+\text{second-order terms},\\
\frac12n_3&=(\mu_3-\mu)(v^2+2vH+H^2+\theta^2)+(\mu_3+\mu)\he\varphi\varphi.
\end{split}
\end{equation}
Observe that when $\vec\mu=\vec0$, the isospin density, $n_3=-2\mu v^2$, appears as soon as the field condenses. For $M^2>0$, this happens once the chemical potential $\mu$ exceeds the mass $M$. This is the Bose--Einstein condensation.

If we now want to make the ground state $\gr{SU(2)}$-neutral, the first terms in the expressions for charge densities lead to the conditions
\begin{equation}
\mu_1=\mu_2=0,\qquad 
\mu_3=\mu\quad\text{(neutral ground state)}.
\label{vacuum_chempot}
\end{equation}
The static bilinear (mass) part of the Lagrangian~\eqref{Lagrangian_expanded} reads, up to an overall minus sign,
\begin{align}
V_{\text{bilin}}(\phi)=&2v(M^2+2\lambda v^2)H+(M^2+6\lambda v^2)H^2\\
\notag
&+(M^2+2\lambda v^2)\theta^2+(M^2+2\lambda v^2-4\mu^2)\he\varphi\varphi.
\end{align}
From the linear term we see that when charge neutrality is imposed, spontaneous symmetry breaking can only occur for $M^2<0$, that is, when it already appears in the vacuum. This is natural: normally, Bose--Einstein condensation (at zero temperature) sets when charge density starts to be nonzero, but here we keep the isospin density equal to zero, which prevents the field from condensing.

Substituting the vacuum expectation value, $v^2=-M^2/2\lambda$, we find that the Higgs mode has the mass term $4\lambda v^2H^2$, while the phase fluctuation of the condensate, $\theta$, becomes massless. However, the same is not true for $\varphi$. This observation led in the context of color superconductivity to the conclusion that the chemical potential needed to render the so-called 2SC phase color-neutral breaks some of the generators explicitly and the associated NG bosons thus naturally acquire nonzero mass~\cite{He:2005mp,*Ebert:2005fi,*Ebert:2006bq}. In our case this corresponds to the $\mu_3$ chemical potential ``explicitly breaking'' the $\tau_{1,2}$ generators, i.e., giving mass to the $\varphi$ excitation. However, the situation is even worse: the mass squared of $\varphi$ is negative! A similar problem appears in the 2SC phase, as will be shown in the following section. In fact, the presence of terms in the Lagrangian~\eqref{Lagrangian_expanded} with a single time derivative makes the discussion slightly more complicated than just concluding that $\varphi$ has a negative mass squared. One finds that the (anti)particle mode annihilated by $\varphi$ ($\he\varphi$) has the dispersion $E_{\vek k}=|\vek k|{\mp}2\mu$. Consequently, the particle mode with momentum smaller than $2\mu$ is ``tachyonic''. To conclude, fixing the chemical potential to make the ground state neutral as we just did obviously leads to an unphysical result.

The roots of the problem lie in the fact that we fixed the chemical potentials in the ground state once for all. The fluctuations of the order parameter then drive the system off the neutrality, thus naturally lowering the energy. This feature was already observed in Ref.~\cite{He:2005jq,*Blaschke:2005km}, where it was erroneously interpreted as an instability of the 2SC ground state. Buballa and Shovkovy~\cite{Buballa:2005bv} pointed out that the instability disappears when the chemical potentials necessary to make the ground state neutral are transformed simultaneously with the ground state itself. It should be stressed that our approach is technically very close to that of Ref.~\cite{Buballa:2005bv}: after all, the NG collective modes in the infinite-wavelength limit correspond merely to a change of the ground state. However, we make one step further by demanding charge neutrality also for certain non-equilibrium field configurations, corresponding to collective modes in the infinite-wavelength limit. In other words, we will require that for every uniform field configuration the chemical potentials acquire such values that the system has zero $\gr{SU(2)}$ charge. This is reasonable since otherwise there would be a uniform charge distribution, leading to an ill-defined thermodynamic limit. Such considerations are sufficient to determine the mass spectrum of the theory, and we will now demonstrate that the NG bosons are indeed exactly massless as they should. A formalism how to deal with nonuniform field fluctuations, that allows us to determine the NG boson dispersion relations, will be developed below.

Let us therefore assume constant fields $H,\theta,\varphi$ and demand that the $\gr{SU(2)}$ charge densities~\eqref{charge_density} are zero for all values of the fields. This gives the chemical potentials the following values,
\begin{equation}
\begin{split}
\mu_1&=-\frac{2\mu}v\,\mathrm{Re}\,\varphi,\qquad\mu_2=+\frac{2\mu}v\,\mathrm{Im}\,\varphi,\\
\mu_3&=\mu-\frac{2\mu}{v^2}\he\varphi\varphi\qquad\text{(neutral excitations)},
\end{split}
\end{equation}
to lowest nontrivial order in the fields. Substituting this back into Eq.~\eqref{Lagrangian_expanded}, the term $\vec\mu\cdot\vec n$ vanishes by definition of charge neutrality, while the rest reduces to
\begin{equation}
\begin{split}
V_{\text{bilin}}(\phi)=&2v(M^2+2\lambda v^2)H+(M^2+6\lambda v^2)H^2\\
&+(M^2+2\lambda v^2)\theta^2+(M^2+2\lambda v^2)\he\varphi\varphi.
\end{split}
\end{equation}
Upon minimizing the potential to make the linear term vanish, the field $\varphi$ becomes exactly massless as it should.


\subsection{Effective field theory approach}
\label{Subsec:EFT}

The masslessness of both $\theta$ and $\varphi$ can be achieved naturally once the chemical potentials are treated as dynamical fields and the local symmetry transformation which generates the NG excitations is accompanied by a corresponding transformation of the chemical potentials. The validity of the Goldstone theorem is then an immediate consequence of the exact symmetry of the action. Let us explain on our simple example the line of reasoning. In the underlying gauge theory the charge neutrality is maintained by a gluon condensate which compensates for the charge of the primary order parameter, here the vacuum expectation value $\phi_0$. Under a symmetry transformation, the primary order parameter and the gluon condensate transform simultaneously. In order to correctly capture the symmetry properties of the gauge theory in a model with a global symmetry, we must allow the chemical potentials to transform in the same way the gluon condensate would. 

Technically this means that we deal with a theory with a global symmetry and two order parameters: the primary $\phi_0$, and the secondary one, that is, the chemical potentials. A low-energy effective Lagrangian for the NG bosons is then constructed as usual by performing a spacetime-dependent symmetry transformation on the order parameter(s). In our specific example, let us parameterize the scalar field as $\phi=\mathcal U(\pi)\phi_0$, where $\mathcal U(\pi)=\exp(\frac iv\vec\tau\cdot\vec\pi)$. Analogously, the temporal background gauge field $\mathcal A$ that contains the chemical potentials and enters via the covariant derivative, $\mathcal D_\mu\phi=(\partial_\mu-i\delta_{\mu0}\mathcal A)\phi$, is parameterized as
\begin{equation}
\mathcal A=\mathcal U(\pi)\mathcal A_0\he{\mathcal U(\pi)},
\label{adjoint}
\end{equation}
where $\mathcal A_0=\mu+\vec\tau\cdot\vec\mu$ with the ground state values of the chemical potentials determined by Eq.~\eqref{vacuum_chempot}. Note that this parameterization abandons the amplitude mode $H$, that is, it defines the nonlinear sigma model. 

Inserting the parameterization into the Lagrangian~\eqref{Lagrangian}, one finds
\begin{equation}
\begin{split}
\mathcal L=&\he\phi_0(\partial_\mu\he{\mathcal U})(\partial^\mu\mathcal U)\phi_0+i\he\phi_0\mathcal A_0(\he{\mathcal U}\partial_0\mathcal U)\phi_0\\
&-i\he\phi_0(\partial_0\he{\mathcal U}\,\mathcal U)\mathcal A_0\phi_0+\he\phi_0\mathcal A_0\mathcal A_0\phi_0\\
&-M^2\he\phi_0\phi_0-\lambda(\he\phi_0\phi_0)^2.
\end{split}
\end{equation}
It is easy to see that the second, third, and fourth term drops because $\mathcal A_0\phi_0=0$, while the last two terms do not include the NG fields $\vec\pi$. The Lagrangian therefore reduces to $\mathcal L=\he\phi_0(\partial_\mu\he{\mathcal U})(\partial^\mu\mathcal U)\phi_0$ up to a constant. Interestingly, it does not depend on the chemical potentials at all. In particular we can see that all NG bosons $\vec\pi$ are exactly massless and have the usual Lorentz-invariant dispersion relation. This is due to the fact that despite the chemical potentials, the densities of all conserved charges, including the $\gr{U(1)}$ charge with respect to which neutrality is not required, are zero in the ground state~\footnote{This is an artifact of the scalar toy model we consider here. The ground state is automatically neutral under an unbroken $\gr{U(1)}'$ group generated by a linear combination of the $\gr{U(1)}$ charge and one of the $\gr{SU(2)}$ charges. Together with the imposed $\gr{SU(2)}$ neutrality, this is already sufficient to guarantee vanishing of all four conserved charges of the $\gr{SU(2)\times U(1)}$ group in the ground state.}.


\section{NG boson masses in NJL model}
\label{Sec:NGmasses} 

The aim of this section is to show that when color neutrality is implemented using the strategy outlined above, the NG bosons become exactly massless in the NJL model description of color-superconducting quark matter. Our plan is as follows. We first introduce the model and essentially repeat the calculation of Ref.~\cite{He:2005mp,*Ebert:2005fi,*Ebert:2006bq} for fixed color chemical potentials. Next, we adapt the approach introduced in Section~\ref{Subsec:instability} for the present purposes. Finally, in Section~\ref{Subsec:corrections} we present some details of the computation of the corrected mass spectrum of the NG bosons.

For the sake of simplicity, we will use a version of the model considered in Ref.~\cite{He:2005mp,*Ebert:2005fi,*Ebert:2006bq} which does not take into account dynamically generated constituent quark masses. This cannot change the conclusions qualitatively. Moreover, in color superconductors the constituent masses of $u$ and $d$ quarks are usually small. Since the calculation is rather technical, we will omit details, yet providing the key steps and definitions necessary for the reader who might desire to reproduce our results.

The model is defined by the Lagrangian in Minkowski space,
\begin{equation}
\mathcal L=\bar q(i\slashed\partial+\gamma_0\mu_\alpha T_\alpha-m)q+
\frac G4\sum_a({\bar q}^{\mathcal C}P_aq)(\bar q\bar P_aq^{\mathcal C}),
\end{equation}
where $q^{\mathcal C}\equiv C\bar q^T$ is the charge conjugated Dirac spinor and the matrices $P_a$ specify the structure of Cooper pairs. In the 2SC phase, they read $(P_a)^{ij}_{bc}=\gamma_5\epsilon^{ij}\epsilon_{abc}$, where $i,j$ are flavor and $a,b,c$ fundamental color indices, respectively. Also, $\bar P_a=\gamma_0\he P_a\gamma_0$. We introduced nine chemical potentials $\mu_\alpha$, $\alpha=0,\dotsc,8$, associated with the symmetry generators $T_\alpha$ in the color space, normalized by $\tr(T_\alpha T_\beta)=\frac12\delta_{\alpha\beta}$. They include the quark number chemical potential $\mu$, related to $\mu_0$ by $\mu_0=\mu\Sqrt6$ since $T_0=\openone/\Sqrt6$.

In the mean-field approximation, the thermodynamic potential $\Omega$ of the model is given by
\begin{equation}
\beta\Omega=\beta V\frac{\Delta_a\Delta^*_a}G-\frac12\mathrm{Tr}\log S^{-1},
\label{NJL_TD}
\end{equation}
where $V$ is the space volume and $\Delta_a$ is the collective field that stands for the composite operator $\frac G2{\bar q}^{\mathcal C}P_aq$. Its ground state expectation value is determined by the minimization of the thermodynamic potential. Furthermore, the quark propagator in the Nambu space, $\Psi\equiv(q,q^{\mathcal C})^T$, reads
\begin{equation}
S^{-1}(k)=\begin{pmatrix}
\slashed k+\gamma_0\mu_\alpha T_\alpha-m & \Delta_a\bar P_a\\
\Delta^*_aP_a & \slashed k-\gamma_0\mu_\alpha T_\alpha^T-m
\end{pmatrix}.
\label{Nambu_propagator}
\end{equation}
The symbol ``$\mathrm{Tr}$'' in Eq.~\eqref{NJL_TD} denotes a trace in the operator sense.

Let us introduce some further notation. In the following, we will not need to work with the general orientation of the condensate $\Delta_a$ in the color space as well as with all the chemical potentials $\mu_\alpha$. In fact, all integrals will be evaluated with only the $\Delta_3\equiv\Delta$ and $\mu_{0,8}$ components nonzero. The quasiquark spectrum is then easily determined analytically, and the dispersion relations of the individual fundamental colors, denoted as $r,g,b$, are $E^e_{\vek k(r,g)}=\Sqrt{(\xi^e_{\vek k(r,g)})^2+\Delta^2}$, $E^e_{\vek k(b)}=|\xi^e_{\vek k(b)}|$, $e=\pm$, with
\begin{equation}
\begin{split}
\xi^e_{\vek k(r)}&=\xi^e_{\vek k(g)}=\epsilon_{\vek k}+e\Bigl(\mu+\frac{\mu_8}{2\Sqrt3}\Bigr),\\
\xi^e_{\vek k(b)}&=\epsilon_{\vek k}+e\Bigl(\mu-\frac{\mu_8}{\Sqrt3}\Bigr),\qquad
\epsilon_{\vek k}=\Sqrt{\vek k^2+m^2}.
\end{split}
\end{equation}
At zero temperature to which we will from now on limit our discussion, the density of an individual (fundamental) quark color $a$ is
\begin{equation}
n_{a}=2\sum_{e=\pm}\int\frac{d^3\vek k}{(2\pi)^3} \frac{e\xi^e_{\vek
k(a)}}{E^e_{\vek k(a)}}.
\label{color_density}
\end{equation}


\subsection{Calculation for fixed chemical potentials}
\label{Subsec:fixedmu}

With the particular orientation of the condensate that we chose, only the $T_8$ generator of the color $\gr{SU(3)}$ develops nonzero density, $n_8=(n_{r}+n_{g}-2n_{b})/(2\Sqrt3)$. It can be made to vanish by tuning $\mu_8$ appropriately. Let us assume that this has been done and calculate the propagator of the four modes that couple to $\Delta_{1,2}$. These form a complex doublet under the unbroken $\gr{SU(2)}$ subgroup and, had we not imposed the neutrality constraint, they would give rise to two type-II NG bosons with quadratic dispersion relation at low momentum~\cite{Nielsen:1975hm,Blaschke:2004cs}. A necessary condition for these to appear is nonzero density of some of the color charges (see Sch\"afer \emph{et al.} in Ref.~\cite{Schaefer:2001bq}). That is why one expects to recover four usual (type-I) NG bosons once the system is made neutral.

Instead, we will show that once the color chemical potential is adjusted in order to make the system color neutral, the propagator of $\Delta_{1,2}$ has a double pole at the frequency $\omega_0=\mu_8\Sqrt3/2$. This was in Ref.~\cite{He:2005mp,*Ebert:2005fi,*Ebert:2006bq} misinterpreted as a manifestation of two pseudo-NG states with degenerate masses. On the contrary, it actually means that the system exhibits an instability of the same type as revealed in the scalar toy model in Section~\ref{Subsec:instability}. In order to understand this, one should note that the 2SC phase possesses an unbroken $\gr{U(1)}_{\tilde{B}}$ symmetry, corresponding to the blue quark number and generated by $\tilde{B}=B-2T_8/\Sqrt{3}$ with $B$ the baryon number. In the space of $\Delta_a$, which transforms in the antitriplet representation of $\gr{SU(3)}$, this symmetry is generated by the matrix $\mathrm{diag}(1,1,0)$. The two degrees of freedom contained in the complex field $\Delta_1$ (and equivalently $\Delta_2$) carry opposite charges, $\tilde{B}=\pm1$; they are a particle--antiparticle pair. Recalling the Umezawa--Kamefuchi--K\"all\'en--Lehmann spectral representation, the particle pole should show up at positive frequencies, while the antiparticle one at negative frequencies. Since the chemical potential required to make the 2SC state neutral is typically negative, the double pole at $\omega_0$ thus actually describes an antiparticle with mass $|\omega_0|$ and a particle with a negative mass, $-|\omega_0|$. This is yet another manifestation of the seeming instability of the color-neutral 2SC state with respect to fluctuations that generate off-diagonal color charges~\cite{He:2005jq,*Blaschke:2005km}. In the following, we will show how this problem can be fixed in a way that renders the theory stable and the NG bosons exactly massless. 

Let us now proceed to the proof of the existence of the double pole in the $\Delta_{1}$ propagator. This is defined as usual by $\Pi_{11}^{-1}(x-y)=-i\langle0|T\{\Delta_1(x)\Delta_1^*(y)\}|0\rangle$, where ``$T$'' denotes time ordering. The inverse propagator, or polarization function, $\Pi_{11}$ is most conveniently evaluated in the random phase approximation. Upon performing Fourier transformation to momentum space and setting the momentum to zero, one finds
\begin{equation}
\begin{split}
\Pi_{11}(\omega)=&\frac1G-2\sum_{e=\pm}\int\frac{d^3\vek k}{(2\pi)^3}\frac1{E^e_{\vek k(r)}}\Biggl[\theta(e\xi^e_{\vek k(b)})\\
&\times\frac{E^e_{\vek k(r)}+e\xi^e_{\vek k(r)}}{\omega+E^e_{\vek k(r)}+|\xi^e_{\vek k(b)}|}\\
&-\theta(-e\xi^e_{\vek k(r)})\frac{E^e_{\vek k(r)}-e\xi^e_{\vek k(r)}}{\omega-E^e_{\vek k(r)}-|\xi^e_{\vek k(b)}|}\Biggr].
\end{split}
\label{invprop}
\end{equation}
The polarization function $\Pi_{22}$ is identical. Using the gap equation at zero temperature,
\begin{equation}
\frac1G=2\sum_{e=\pm}\int\frac{d^3\vek k}{(2\pi)^3}\frac1{E^e_{\vek k(r)}},
\end{equation}
the inverse propagator~\eqref{invprop} is easily brought to the form
\begin{equation}
\begin{split}
&\Pi_{11}(\omega)=(2\omega-\mu_8\Sqrt3)\sum_{e=\pm}\int\frac{d^3\vek k}{(2\pi)^3}\frac1{E^e_{\vek k(r)}}\\
&\times\left[\frac{\theta(e\xi^e_{\vek k(b)})}{\omega+E^e_{\vek k(r)}+|\xi^e_{\vek k(b)}|}+\frac{\theta(-e\xi^e_{\vek k(b)})}{\omega-E^e_{\vek k(r)}-|\xi^e_{\vek k(b)}|}\right].
\end{split}
\label{doublepole}
\end{equation}

The prefactor in Eq.~\eqref{doublepole} immediately tells us that there is a pole at $\omega=\omega_0$. Since this pole becomes exactly massless in the limit of $\mu_8=0$, thereby representing a NG boson, it is the particle pole. To find the antiparticle pole requires some further manipulation. Evaluating the expression in brackets at $\omega=\omega_0$ yields
\begin{equation}
\begin{split}
&\frac{\theta(e\xi^e_{\vek k(b)})}{\frac{\mu_8\Sqrt3}2+E^e_{\vek k(r)}+e\xi^e_{\vek k(b)}}+\frac{\theta(-e\xi^e_{\vek k(b)})}{\frac{\mu_8\Sqrt3}2-E^e_{\vek k(r)}+e\xi^e_{\vek k(b)}}\\
&=\frac{\theta(e\xi^e_{\vek k(b)})}{E^e_{\vek k(r)}+e\xi^e_{\vek k(r)}}+\frac{\theta(-e\xi^e_{\vek k(b)})}{-E^e_{\vek k(r)}+e\xi^e_{\vek k(r)}}\\
&=\frac{E^e_{\vek k(r)}\,\mathrm{sgn}(e\xi^e_{\vek k(b)})-e\xi^e_{\vek k(r)}}{\Delta^2}.
\end{split}
\label{manipulation}
\end{equation}
Using Eq.~\eqref{color_density}, one then arrives at the conclusion that the integral (including the sum over $e$) in Eq.~\eqref{doublepole} equals $-n_8\Sqrt3/(2\Delta^2)$. In fact, it was already observed in Ref.~\cite{Brauner:2008td} that the coefficient of the term linear in $\omega$ in the expansion of the inverse propagator around the particle pole is proportional to the density $n_8$. The present result is just a generalization to nonzero values of $\mu_8$. We therefore conclude that when the chemical potential $\mu_8$ is tuned so that $n_8=0$, the propagator indeed has a double pole at $\omega=\omega_0$.

We note in passing that, as the second line of Eq.~\eqref{manipulation} clearly shows, the double pole occurs outside the two-body continuum, so the implied instability cannot be alleviated by decay processes.


\subsection{Induced fluctuations of chemical potentials}
\label{Subsec:induced}

We would now like to apply the same strategy as in Section~\ref{Subsec:instability} to remove the instability revealed above, and to show that all NG modes in the 2SC phase are exactly massless as they should. However, due to the complicated form of the mean-field thermodynamic potential in the NJL model, it is not possible to solve analytically for the color chemical potentials as a function of the (uniform) collective fields. Fortunately, this is not really necessary if we are only interested in the mass spectrum of the collective excitations.

Let us consider generally a system possessing a set of (real) order parameters, $\Delta_a$, constrained by the requirement of zero densities, $n_\alpha$, of a set of conserved charges. Introducing the associated chemical potentials, $\mu_\alpha$, its thermodynamics is governed by the grand canonical potential, $\Omega(\Delta_a,\mu_\alpha)$. The values of the order parameters and the chemical potentials in the ground state are determined by the set of gap equations, $\partial\Omega/\partial\Delta_a=0$, together with the constraints,
\begin{equation}
\frac{\partial\Omega}{\partial\mu_\alpha}=0.
\label{constraint}
\end{equation}
In order to study the fluctuations in a neutral system, one uses this equation to eliminate the chemical potentials in favor of the order parameters, and thereby arrives at a thermodynamic potential as a function of $\Delta_a$ solely, $\tilde\Omega\bigl(\Delta_a,\mu_\alpha(\Delta_a)\bigr)$. Such a thermodynamic potential has a minimum that coincides with the simultaneous solution of the gap and constraint equations for $\Omega(\Delta_a,\mu_\alpha)$. In an unconstrained system, the mass matrix of the collective modes is, up to an irrelevant factor, proportional to $\partial^2\Omega/\partial\Delta_a\partial\Delta_b$. Once the constraint~\eqref{constraint} is imposed, this must be obviously replaced with the total second derivative, $d^2\tilde\Omega/d\Delta_ad\Delta_b$. Differentiating Eq.~\eqref{constraint} with respect to $\Delta_a$ (this builds in the requirement that the constraint be satisfied for all values of the order parameter, at least in the vicinity of the equilibrium), one arrives at
\begin{align}
\label{total_mass}
\frac{d^2\tilde\Omega}{d\Delta_ad\Delta_b}&=\frac{\partial^2\Omega}{\partial\Delta_a\partial\Delta_b}-\frac{\partial^2\Omega}{\partial\mu_\alpha\partial\mu_\beta}\frac{\partial\mu_\alpha}{\partial\Delta_a}\frac{\partial\mu_\beta}{\partial\Delta_b}\\
\notag
&=\frac{\partial^2\Omega}{\partial\Delta_a\partial\Delta_b}-\frac{\partial^2\Omega}{\partial\Delta_a\partial\mu_\alpha}\left(\frac{\partial^2\Omega}{\partial\mu\partial\mu}\right)^{-1}_{\alpha\beta}\frac{\partial^2\Omega}{\partial\Delta_b\partial\mu_\beta}.
\end{align}
All partial derivatives are to be evaluated at the ground state values of the order parameters and chemical potentials.

In the next subsection it will be demonstrated explicitly that this prescription results in exactly massless NG bosons in the color-neutral 2SC phase. Here we just note that this is very natural: the potential $\Omega(\Delta_a,\mu_\alpha)$ is invariant under simultaneous transformations of the order parameters and the chemical potentials. Therefore, the potential $\tilde\Omega\bigl(\Delta_a,\mu_\alpha(\Delta_a)\bigr)$ is invariant under the symmetry transformation of $\Delta_a$, and the masslessness of NG bosons follows immediately from the Goldstone theorem.

Before proceeding to the explicit NJL-model calculation, we would like to point out one subtlety hidden in Eq.~\eqref{total_mass}. The matrix $\partial^2\Omega/\partial\mu_\alpha\partial\mu_\beta$ in Eq.~\eqref{total_mass} represents minus the density--density correlator (or the color number susceptibility matrix) in the ground state so that it is negative-semidefinite. The charges of the unbroken symmetry have by definition sharp values in the ground state, and thus give rise to zero modes of the correlator. The matrix of second partial derivatives is therefore not invertible. There is a simple remedy: one adds a term $-\zeta^2\mu_\alpha\mu_\alpha$ to the thermodynamic potential which makes all expressions well defined. In the end, the limit $\zeta\to0$ is performed. Since the unbroken generators do not couple to NG bosons, this subtlety does not affect the calculation of their mass spectrum.


\subsection{Induced corrections to NG boson masses}
\label{Subsec:corrections}

We will again focus on the complex doublet of NG bosons that couple to $\Delta_{1,2}$~\footnote{There is one more NG boson annihilated by $\Delta_3$, associated with the spontaneous breaking of the $T_8$ generator. Its masslessness is not questioned because $T_8$ is not ``explicitly broken'' by the chemical potential $\mu_8$.}. Recall that these fields are complex and we are interested in the derivative $d^2\tilde\Omega/d\Delta_1^*d\Delta_1$ which yields the static part of the inverse propagator~\eqref{invprop}.

We need the three second partial derivatives of the thermodynamic potential, $\partial^2\Omega/\partial\Delta^*_1\partial\Delta_1,$ $\partial^2\Omega/\partial\mu_\alpha\partial\mu_\beta$, and $\partial^2\Omega/\partial\Delta^*_1\partial\mu_\alpha$. The first one is already contained in Eq.~\eqref{doublepole},
\begin{equation}
\begin{split}
\frac1V\frac{\partial^2\Omega}{\partial\Delta^*_1\partial\Delta_1}&=-\mu_8\Sqrt3\,X,\\
X&=\sum_{e=\pm}\int\frac{d^3\vek k}{(2\pi)^3}\frac1{E^e_{\vek k(r)}} \frac{\mathrm{sgn}(e\xi^e_{\vek k(b)})}{E^e_{\vek k(r)}+|\xi^e_{\vek k(b)}|}.
\end{split}
\label{piece1}
\end{equation}
The density--density correlator is given by two one-loop diagrams, with normal and anomalous components of the quark propagator. Carrying out the partial derivative of the thermodynamic potential~\eqref{NJL_TD}, we obtain
\begin{equation}
\beta\frac{\partial^2\Omega}{\partial\mu_\alpha\partial\mu_\beta}=\mathrm{Tr}\left(\gamma_0T_\alpha S_{q\bar q}\gamma_0T_\beta S_{q\bar q}-\gamma_0T_\alpha S_{qq}\gamma_0T_\beta^TS_{\bar q\bar q}\right).
\label{auxeq0}
\end{equation}
The subscripts of the propagator denote matrix elements in the Nambu space. Inserting the quark propagator~\eqref{Nambu_propagator} and carrying out the frequency integration, the first term evaluates to
\begin{equation}
\begin{split}
&-\frac{\beta V}2\Delta^2\tr(T_\alpha\mathcal P_{12}T_\beta\mathcal P_{12})Y\\
&-\beta V\left[\tr(T_\alpha\mathcal P_{12}T_\beta\mathcal P_3)+\tr(T_\alpha\mathcal P_3T_\beta\mathcal P_{12})\right]Z,
\end{split}
\label{auxeq}
\end{equation}
where
\begin{align}
Y&=\sum_{e=\pm}\int\frac{d^3\vek k}{(2\pi)^3}\frac1{(E^e_{\vek k(r)})^3},\\
\notag
Z&=\sum_{e=\pm}\int\frac{d^3\vek k}{(2\pi)^3}\frac1{E^e_{\vek k(r)}|\xi^e_{\vek k(b)}|}\frac{E^e_{\vek k(r)}|\xi^e_{\vek k(b)}|-\xi^e_{\vek k(r)}\xi^e_{\vek k(b)}}{E^e_{\vek k(r)}+|\xi^e_{\vek k(b)}|},
\end{align}
and $\mathcal P_{12}$ and $\mathcal P_3$ are projectors on the subspace of first two colors and the third color, respectively. The symbol ``$\tr$'' stands for a trace over color and flavor indices here. The second term in Eq.~\eqref{auxeq0} reduces to $-(\beta V/2)\tr(T_\alpha MT_\beta^T\he M)Y$, where $M$ is a matrix in color--flavor space defined by $M=\gamma_5P_3$. Putting all the pieces together, we obtain
\begin{equation}
\begin{split}
&\frac1V\frac{\partial^2\Omega}{\partial\mu_\alpha\partial\mu_\beta}\\
&=-\frac13\begin{pmatrix}
2 & \Sqrt2\\
\Sqrt2 & 1
\end{pmatrix}\Delta^2Y\quad\text{in the $(T_0,T_8)$ sector},\\
&=-Z\delta_{\alpha\beta}\quad\text{for $\alpha=4,5,6,7$},
\end{split}
\label{piece2}
\end{equation}
and zero otherwise. Finally, the mixed partial derivative of $\Omega$ is given by a one-loop diagram with one normal and one anomalous propagator,
\begin{equation}
\beta\frac{\partial^2\Omega}{\partial\Delta_a^*\partial\mu_\alpha}=\mathrm{Tr}\left(\gamma_0T_\alpha S_{qq}P_aS_{q\bar q}\right),
\end{equation}
which, after appropriate manipulations, yields
\begin{align}
\notag
\frac1V\frac{\partial^2\Omega}{\partial\Delta_a^*\partial\mu_\alpha}=&\tr(T_\alpha M\gamma_5P_a\mathcal P_3)\,X+\frac12\tr(T_\alpha M\gamma_5P_a\mathcal P_{12})\\
&\times\sum_{e=\pm} \int\frac{d^3\vek k}{(2\pi)^3}\frac{e\xi^e_{\vek k(r)}}{(E^e_{\vek k(r)})^3}. 
\label{piece3}
\end{align}
For the first component of the diquark field, the color--flavor traces are $\tr(T_\alpha M\gamma_5P_1\mathcal P_{12})=0$ and $\tr(T_\alpha M\gamma_5P_1\mathcal P_3)=(0,0,0,-\Delta,-i\Delta,0,0,0)$ for $\alpha=1,\dotsc,8$.

Finally, we combine Eqs.~\eqref{total_mass}, \eqref{piece1}, \eqref{piece2}, and \eqref{piece3} to arrive at the simple result
\begin{equation}
\frac1V\frac{d^2\Omega}{d\Delta_1^*d\Delta_1}=-\Sqrt3\,\frac XZ\biggl(\mu_8Z
-\frac2{\Sqrt3}\Delta^2X\biggr).
\end{equation}
A slight manipulation with the integrals reveals that the expression in the parentheses equals the charge density $n_8$. This concludes the argument that once Eq.~\eqref{total_mass} is used instead of a simple second partial derivative to calculate the mass spectrum, the NG bosons are rendered exactly massless, as required by the Goldstone theorem.


\section{Dispersion relations of NG bosons in high-density approximation}
\label{Sec:NGdispersions}

The approach taken in the previous section was straightforward, yet intrinsically limited to uniform order parameter fluctuations, so demonstrating the masslessness of the NG modes was the best it could do for us. Here we will follow the path outlined in Section~\ref{Subsec:EFT} and construct the effective action for the NG modes in the 2SC phase. We will use the high-density effective theory (HDET, see Ref.~\cite{Nardulli:2002ma} for a review), providing model-independent results in the limit of high baryon density.

The HDET Lagrangian is constructed using quark fields and their coupling to colored diquarks. For the 2SC superconductor, it takes the following form,
\begin{align}
\notag
\mathcal L_{\text{HDET}}=&\frac{1}{2}\sum_{\vek{v}}\Bigl[\he{q_+}(iV\cdot\partial+\hat{\mu}) q_++\he{q_-}(i\tilde{V}\cdot\partial+\hat{\mu})q_-\\
\label{HDETlag}
&+\Delta_{a}^*q_-^T{\tau}_2{\vek{\epsilon}}_a q_++\Delta_{a}\he{q_+}{\tau}_2\vek{\epsilon}_aq_-^*\Bigr]\\
\notag
&+(\text{L}\to\text{R},\Delta_{a}\to-\Delta_{a}).
\end{align}
Here $q$ stands for the left-handed positive-energy-projected quark field and the subscript $\pm$ denotes the (conjugated) velocity. The two four-velocities appearing in the Lagrangian are defined as $V^\mu=(1,\vek{v})$ and $\tilde{V}^\mu=(1,-\vek{v})$; the sum is taken over pairs of patches on the Fermi surface characterized by velocities $\pm\vek v$. Furthermore, ${\tau}_2$ is the Pauli matrix in the flavor space and $\hat{\mu}$ is the chemical potential matrix in the color space; we consider the form $\hat{\mu}=\mu\openone+\sum_{\alpha=1}^8\mu_\alpha T_\alpha$ ignoring the electric charge neutrality. Also $\vek{\epsilon}_a=2T_7,2T_5,2T_2$ for $a=1,2,3$ are the color antisymmetric generators, and accordingly the $\Delta_{a}$'s stand for the background quark pair fields. The subscript $a$ represents the color antitriplet indices which can be thought of as color-antisymmetric pairs, $[g,b],[b,r],[r,g]$, respectively. Finally, the ``$\mathrm{L\to R}$'' in the last line of Eq.~\eqref{HDETlag} is to remind us that an equivalent Lagrangian for the right-handed quarks has to be added.

By introducing the Nambu doublet field, $\Psi\equiv(q_+,-q_-^*)^T$, the Lagrangian~\eqref{HDETlag} becomes
\begin{equation}
\begin{split}
\mathcal L_{\text{HDET}}=&\frac{1}{2}\sum_{\vek{v}}\he\Psi
\begin{pmatrix}
iV\cdot\partial+\hat{\mu} & -\Delta_a\tau_2\vek\epsilon_a\\
-\Delta^*_a\tau_2\vek\epsilon_a & i\tilde{V}\cdot\partial-\hat{\mu}^*
\end{pmatrix}\Psi\\
&+(\text{L}\to\text{R},\Delta_a\to-\Delta_a).
\end{split}
\label{HDETefflag}
\end{equation}
Let us construct the effective Lagrangian for the NG modes. Following the logic explained in Section~\ref{Sec:toymodel} we have to consider fluctuations of both the pairing field $\Delta_a$ and the color chemical potential matrix $\hat\mu$, which is treated as a secondary order parameter induced by the color neutrality constraint. Without loss of generality, we adopt the convention that the ground state is characterized by the nonzero condensate $\Delta_3=\Delta_3^*\equiv\Delta$. Accordingly, only the color chemical potential in the $T_8$-direction is nonzero in the ground state. The actual values of $\Delta$ and $\mu_8$ are set by the gap equation and the neutrality constraint. Having all this in mind, we parameterize the NG fields $\Delta_a$ and chemical potentials $\mu_\alpha$ in Eq.~\eqref{HDETefflag} as
\begin{equation}
\begin{split}
\Delta_a\vek\epsilon_a&=\mathcal U(\vek{\pi})2\Delta T_2\mathcal U(\vek{\pi})^T,\\
\hat\mu&=\mu\openone+\mathcal U(\vek{\pi})\mu_8T_8\he{\mathcal U(\vek{\pi})},
\end{split}
\label{eq:parametrization}
\end{equation}
where $\mathcal U$ is expressed in terms of the NG fields $\vek{\pi}$ as
\begin{equation}
\mathcal U(\vek{\pi})=\exp\biggl[\frac i{f_\pi}\sum_{\alpha=4}^8\pi_\alpha(x)T_\alpha\biggr],
\label{eq:calU}
\end{equation}
and we introduced the decay constant $f_\pi$. In principle, an
independent decay constant should be used for every real irreducible representation of the unbroken global symmetry, in this case one for $\alpha=4,5,6,7$ and one for $\alpha=8$. We just use the same symbol for them in order to simplify the notation.

It is now easy to obtain an effective action for the NG modes in a gradient expansion. Following essentially the same steps as in Section~\ref{Sec:NGmasses}, we integrate out the quark fields to obtain
\begin{equation}
S_{\text{eff}}=-\mathrm{Tr}\log S^{-1},
\end{equation}
where $S$ is the quark propagator in Nambu space in presence of the fluctuating order parameters. This formula is to be contrasted to Eq.~\eqref{NJL_TD}. The mass term $\Delta_a\Delta_a^*/G$ is missing here and consequently the NG nature of the order parameter fluctuations is made manifest.

We find the following explicit expression for the propagator up to the second order in the NG fields,
\begin{equation}
\begin{split}
S^{-1}(x,y)=&\begin{pmatrix}
iV\cdot\partial+\hat{\mu} & -2\Delta\tau_2T_2\\
-2\Delta\tau_2T_2 & i\tilde{V}\cdot\partial-\hat{\mu}
\end{pmatrix}
\delta^4(x-y)\\
&+\frac{\pi_\alpha(x)}{f_\pi}\Sigma^{(1)}_{\alpha}\delta^4(x-y)\\
&-\frac{\pi_\alpha(x)\pi_\beta(x)}{f_\pi^2}\Sigma^{(2)}_{\alpha\beta}\delta^4(x-y),
\label{propagator}
\end{split}
\end{equation}
where
\begin{equation}
\begin{split}
\Sigma^{(1)}_\alpha&=\begin{pmatrix}
-\mu_8\Xi_\alpha&\Delta\tau_2\Gamma_\alpha\\
\Delta\tau_2\he{\Gamma_\alpha}&+\mu_8\Xi_{\alpha}^*
\end{pmatrix},\\
\Sigma^{(2)}_{\alpha\beta}&=\begin{pmatrix}
-\mu_8\Xi_{\alpha\beta}&\Delta\tau_2\Gamma_{\alpha\beta}\\
\Delta\tau_2\he{\Gamma_{\alpha\beta}}&+\mu_8\Xi_{\alpha\beta}^*
\end{pmatrix},
\end{split}
\end{equation}
and $\Sigma_{\alpha}$, $\Gamma_\alpha$, $\Sigma_{\alpha\beta}$, and $\Xi_{\alpha\beta}$ are matrices in the fundamental color space. The explicit forms of these matrices are found by expanding Eq.~\eqref{eq:parametrization} up to second order in the NG fields. We note that the diagonal entries proportional to $\mu_8$ take into account the effect of fluctuations of the chemical potentials that was missed in previous analyses~\cite{Blaschke:2004cs,He:2005mp}. The gradient expansion of the action thus reads
\begin{align}
\notag
S_{\text{eff}}=&-\mathrm{Tr}\log S^{-1}_0\\
\notag
&+\int d^4x\frac{\pi_\alpha(x)\pi_\beta(x)}{f_\pi^2}\tr\left[S_0(x,x)\Sigma^{(2)}_{\alpha\beta}\right]\\
&+\int d^4x\,d^4y\frac{\pi_\alpha(x)\pi_\beta(y)}{2f_\pi^2}\\
\notag
&\times\tr\left[S_0(x,y)\Sigma^{(1)}_\alpha S_0(y,x)\Sigma^{(1)}_\beta\right],
\end{align}
with $S_0^{-1}$ being the first term of Eq.~\eqref{propagator}. The first term above is proportional to the thermodynamic potential in the equilibrium. The second term gives the tadpole contribution stemming from the fact that we employed a non-linear parameterization of the collective fields. Omitting the part of zeroth order in the NG fields and working at nonzero temperature within the Matsubara formalism, the bilinear part of the effective action becomes, in momentum space,
\begin{equation}
\begin{split}
S_{\text{eff}}=&\frac{T}{2}\sum_{N}\int\frac{d^3\vek{P}}{(2\pi)^3}\pi_\alpha(-i\Omega_N,-\vek{P})\\
&\times\Pi_{\alpha\beta}(i\Omega_N,\vek{P})\pi_{\beta}(i\Omega_N,\vek{P}),
\end{split}
\end{equation}
where $\Omega_N=2N\pi T$ and $N$ is an integer. Reality of the NG fields in the coordinate space implies the constraint on their Fourier components, $\pi_{\alpha}(-i\Omega_N,-\vek{P})=\pi_{\alpha}(i\Omega_N,\vek{P})$. The inverse propagator of the NG fields is given by
\begin{align}
\notag
{\Pi}_{\alpha\beta}(i\Omega_N,\vek{P})=&\frac{T}{f_\pi^2}\sum_{n}\int\frac{d^3\vek{q}}{(2\pi)^3}\tr\left[2S_0(i\omega_n,\vek{q})\Sigma_{\alpha\beta}^{(2)}\right]\\
\label{eq:selfenergy3}
&+\frac{T}{f_\pi^2}\sum_{n}\int\frac{d^3\vek{q}}{(2\pi)^3}\tr\Bigl[S_0(i\omega_n,\vek{q})\\
\notag
&\times\Sigma_{\alpha}^{(1)}S_0(i\omega_n+i\Omega_N,\vek{q}+\vek{P})\Sigma_{\beta}^{(1)}\Bigr],
\end{align}
with $\omega_n=(2n+1)\pi T$ being the fermionic Matsubara frequency. In the Nambu space, the fermion propagator takes the form
\begin{equation}
S_0(i\omega_n,\vek{q})=\begin{pmatrix}
S_{q\bar{q}}(i\omega_n,\vek{q})& S_{qq}(i\omega_n,\vek{q})\\
S_{\bar{q}\bar{q}}(i\omega_n,\vek{q})&S_{\bar{q}q}(i\omega_n,\vek{q})
\end{pmatrix},
\end{equation}
where the individual elements are given explicitly by
\begin{equation}
\begin{split}
S_{q\bar{q}}(i\omega_n,\vek{q})&= (\openone-\tilde{B})S_{q\bar{q}}^{(r,g)}+\tilde{B}S_{q\bar{q}}^{(b)},\\
S_{qq}(i\omega_n,\vek{q})&=2\tau_2T_2S_{qq}^{(r,g)},\\
S_{q\bar{q}}^{(r,g)}&=-\frac{i\omega_n+\vek{v}\cdot\vek{q}-\mu_r}{\omega_n^2+(\vek{v}\cdot\vek{q}-\mu_r)^2+\Delta^2},\\
S_{q\bar{q}}^{(b)}&=\frac{1}{i\omega_n-(\vek{v}\cdot\vek{q}-\mu_b)},\\
S_{qq}^{(r,g)}&=-\frac{\Delta}{\omega_n^2+(\vek{v}\cdot\vek{q}-\mu_r)^2+\Delta^2}.\\
\end{split}
\end{equation}
As before, $\tilde{B}=\frac{1}{3}\openone-2T_8/\Sqrt3=\mathrm{diag}(0,0,1)$ is the projector to the blue quark space, or in other words, the blue quark number in the fundamental color space. The chemical potentials for the red, green, and blue quarks are defined by $\mu_{r,g}=\mu+\mu_8/(2\Sqrt{3})$ and $\mu_b=\mu-\mu_8/\Sqrt{3}$.

Upon carrying out the trace in the color and flavor spaces and performing the Matsubara summation, ${\Pi}_{\alpha\beta}$ turns out to have the following block-diagonal structure in the adjoint color space, dictated by the unbroken symmetry,
\begin{equation}
\begin{split}
 \Pi_{\alpha\beta}=&\begin{pmatrix}
\Pi_{\text{tad}}^{\text{ch}}+\Pi^{\text{ch}} & \Pi_{\text{mix}}\\
-\Pi_{\text{mix}} & \Pi_{\text{tad}}^{\text{ch}}+\Pi^{\text{ch}}\\
\end{pmatrix}_{(\pi_4,\pi_5)}\\
&\oplus
\begin{pmatrix}
\Pi_{\text{tad}}^{\text{ch}}+\Pi^{\text{ch}} & \Pi_{\text{mix}}\\
-\Pi_{\text{mix}} & \Pi_{\text{tad}}^{\text{ch}}+\Pi^{\text{ch}}\\
\end{pmatrix}_{(\pi_6,\pi_7)}\\
&\oplus
(\Pi^{\rm n}_{\text{tad}}+\Pi^{\rm n})_{(\pi_8)}.
\end{split}
\label{eq:polarization}
\end{equation}
The superscripts ``n'' and ``ch'' distinguish modes that are respectively neutral and charged with respect to the unbroken quantum number $\tilde B$. In accord with what we already observed in Section~\ref{Sec:NGmasses}, there should be four charged modes that form a complex doublet of the unbroken global color $\gr{SU(2)}$ symmetry, and one neutral NG mode. Using the unbroken symmetry, the propagator in the charged sector can be diagonalized in the basis $(\pi_4\pm i\pi_5)/\Sqrt2$, $(\pi_6\pm i\pi_7)/\Sqrt2$. The various contributions to the propagator are labelled by subscripts: ``tad'' refers to the tadpole contribution coming from the first term in Eq.~\eqref{eq:selfenergy3}, whereas ``mix'' refers to terms that mix different components $\pi_\alpha$, thereby giving rise to mass splitting of modes with opposite values of $\tilde B$. As we will see below, the spacetime dependence of the propagator at long-wavelengths can be extracted analytically within HDET.


\subsection{NG mode in the neutral sector}

Let us first have a look at the neutral NG mode, $\pi_8$, which corresponds to mere phase fluctuations of the order parameter and hence, based on the toy model analyzed in Section~\ref{Sec:toymodel}, should not suffer from artifacts associated with the color neutrality constraint. Its dispersion relation can be extracted from the gradient expansion of $\Pi^{\rm n}_{\text{tad}}+\Pi^{\rm n}(i\Omega_N,\vek{P})$. Explicit computation yields
\begin{equation}
\Pi^{\rm n}_{\text{tad}}=\frac{4N_{\rm f}}{3f_\pi^2}\Delta\phi(\Delta,\mu_r),
\label{eq:tadneutral}
\end{equation}
with $\phi$ being the anomalous pair density defined by $\langle q_{-}^{ai}q_{+}^{bj}\rangle=-(\tau_2)_{ij}(2T_{2})_{ab}\phi(\Delta,\mu_r)$; $N_{\rm f}=2$ is the number of flavors. In terms of the quark propagator, it is expressed as
\begin{equation}
 \phi(\Delta,\mu_r)=-T\sum_n\int\frac{d^3\vek{q}} {(2\pi)^3}S_{qq}^{(r,g)}(i\omega_n,\vek{q}).
\end{equation}
On the other hand, the gradient expansion of $\Pi^{\rm n}(i\Omega_N,\vek P)$ becomes 
\begin{equation}
\Pi^{\rm n}(i\Omega_N,\vek{P})=\Pi^{\rm n}(0,\vek0)-A(i\Omega_N)^2+B\vek{P}^2+\dotsb,
\end{equation}
with $\Pi^{\rm n}(0,\vek0)=-\Pi^{\rm n}_{\text{tad}}$ which guarantees that the mode is gapless in accord with the Goldstone theorem. The coefficients $A$ and $B$ are evaluated within HDET in the limit of vanishing temperature and at the leading order in $\Delta/\mu,\mu_8/\mu$ as
\begin{equation}
A=\frac{N_{\rm f}N_0}{3f_\pi^2},\qquad
B=\frac{N_{\rm f}N_0}{9f_\pi^2},
\end{equation}
where $N_0\equiv{\mu^2}/{2\pi^2}$ is the density of states at the Fermi surface. In order for the NG mode effective action to take the canonical form with $A=1$, we set the decay constant to $f_\pi^2=N_{\rm f}N_0/3$. The phase velocity of $\pi_8$ becomes
\begin{equation}
v_\pi=\Sqrt{\frac{B}{A}}=\Sqrt{\frac{1}{3}},
\end{equation}
which is the usual result reflecting the number of space dimensions~\cite{Nardulli:2002ma}. Since $\pi_8$ is a singlet under the unbroken global color $\gr{SU(2)}$ symmetry and carries zero $\tilde B$ charge, it is not much affected by the color neutrality constraint. The above conclusion therefore remains correct irrespective of the value of the background color charge density $n_8$ as long as $\mu_8/\mu$ is small. In the following, we will concentrate on the $\tilde{B}$-charged sector of the NG effective action.


\subsection{NG modes in the charged sector}

We are now ready to analyze in detail the excitation spectrum of the charged NG modes in the $(\pi_4,\pi_5)$ sector. [The spectrum in the $(\pi_6,\pi_7)$ sector is identical as a consequence of the unbroken global color $\gr{SU(2)}$ symmetry.] Before going into details, let us briefly pause to overview the general properties of the charged sector. Following the discussion below Eq.~\eqref{eq:polarization}, we define the combinations $\pi_\pm\equiv(\pi_4\pm i\pi_5)/\Sqrt2$ which carry the charges $\tilde B=\pm1$. The low-energy behavior of these can be extracted from the polarization function in the charged $(\pi_4,\pi_5)$-sector in Eq.~\eqref{eq:polarization}. The components of the propagator satisfy the complex conjugation properties $\Pi^{\text{ch}}(-i\Omega_N,\vek{P})=\Pi^{\text{ch}}(i\Omega_N,\vek{P})$ and $\Pi_{\text{mix}}(-i\Omega_N,\vek{P})=-\Pi_{\text{mix}}(i\Omega_N,\vek{P})$. In the basis $(\pi_+,\pi_-)$, the polarization matrix becomes diagonal with the entries $\Pi_\pm=\Pi_{\text{tad}}^{\text{ch}}+\Pi^{\text{ch}}\mp i\Pi_{\text{mix}}$. The charge conjugation then implies that $\Pi_-(i\Omega_N,\vek{P})=\Pi_{+}(-i\Omega_N,\vek{P})$.

In order to clarify the role of the fluctuations of the chemical potentials and the coupling between charge density fluctuations and the NG modes, we decompose the functions $\Pi^{\text{ch}}$ and $\Pi_{\text{mix}}$ as
\begin{equation}
\begin{split}
\Pi^{\text{ch}}&=\Pi_{\Delta\Delta}^{\text{ch}}+\Pi_{\Delta\mu}^{\text{ch}}+\Pi_{\mu\mu}^{\text{ch}},\\
\Pi_{\text{mix}}&=\Pi_{\text{mix}}^{\Delta\Delta}+\Pi_{\text{mix}}^{\Delta\mu}+\Pi_{\text{mix}}^{\mu\mu},
\end{split}
\end{equation}
where the first, second, and third terms are proportional to $\Delta^2$, $\Delta\mu_8$, and $\mu_8^2$, respectively. Diagrammatically, they are expressed as the quark one-loop diagrams containing two anomalous propagators, one normal and one anomalous propagator, and two normal propagators. $\Pi_{\text{tad}}^{\text{ch}}$ is computed explicitly as
\begin{equation}
\Pi_{\text{tad}}^{\text{ch}}=\frac{N_{\text{f}}}{f_\pi^2}\Delta\phi(\Delta,\mu_r)+\frac{3}{4f_\pi^2}\mu_8n_8(\Delta,\hat{\mu}),
\label{eq:tad}
\end{equation}
where $n_8(\Delta,\hat{\mu})\equiv(n_r+n_g-2n_b)/2\Sqrt{3}$ is the color charge density at an arbitrary $\mu_8$. In contrast to the tadpole contribution~\eqref{eq:tadneutral} in the neutral sector, there is an additional term proportional to $\mu_8n_8$ which comes from the tadpole diagram with a quark loop containing the normal propagator. This is because the charged NG mode is sensitive to the secondary order parameter, that is, $\mu_8$. This additional term is absent if we ignore the fluctuations in color chemical potentials.

We remark that $\pi_+,\pi_-$ correspond to $\Delta_1,\Delta_1^*$ in the notation of Section~\ref{Sec:NGmasses}. Therefore, the curvature mass matrix calculated in Section~\ref{Subsec:corrections} can be obtained as the long-wavelength limit of the polarization function,
\begin{equation}
\begin{split}
\frac{1}{V}\frac{d^2\Omega}{d\Delta_1^*d\Delta_1}&=\lim_{\vek{P}\to\vek{0}}\Pi_\pm(0,\vek{P})\\
&=\Pi_{\text{tad}}^{\text{ch}}(0,\vek{0})+\Pi^{\text{ch}}(0,\vek{0}).
\end{split}
\end{equation}

We shall now calculate in detail the polarization functions $\Pi^\pm$, and hence the dispersion relations of the $\pi_\pm$ modes. In order to elucidate the role of the color neutrality and the necessity to implement it carefully, we will analyze three different scenarios:
\begin{itemize}
\item[(i)] The case with $\mu_8=0$. In this case the primary order parameter, that is $\Delta$, induce nonzero color charge density in the system. We expect the spectrum to contain type-II NG bosons.
\item[(ii)] The case of color neutrality ensured by ``hard'' (fixed) chemical potential(s). The chemical potential $\mu_8$ is tuned so that there is no color charge in the ground state, but we ignore the chemical potential fluctuations; this is analogous to the discussion in Section~\ref{Subsec:fixedmu}. Technically this means that we discard the specific contributions to the polarization function which couple to color density such as $\Pi^{\text{ch}}_{\Delta\mu}$, $\Pi^{\text{ch}}_{\mu\mu}$ etc. In this case one may naively expect four NG bosons to acquire nonzero masses since the hard external background $\mu_8$ serves as an explicit symmetry breaking source in the quark sector~\cite{He:2005mp,*Ebert:2005fi,*Ebert:2006bq}.
\item[(iii)] The full analysis of NG bosons in the neutral system. We take into account fluctuations in the chemical potentials and their coupling to NG bosons; this is analogous to the discussion in Section~\ref{Subsec:corrections}. Technically this means including all contributions to the polarization function. We expect in this case to recover five massless type-I NG bosons.
\end{itemize}


\subsubsection{Analysis for the case (i)}

We set $\mu_8=0$ and $\mu_r=\mu_g=\mu_b=\mu$. After analytically continuing the polarization function to real frequencies, $i\Omega_N\to\omega+i\delta$, and expanding in powers of $\omega$ and $\vek{P}$, one finds, up to the second order,
\begin{equation}
\Pi^{\text{ch}}(\omega,\vek{P})=\Pi^{\text{ch}}(0,\vek{0})-A\omega^2+B\vek{P}^2+\dotsb,
\end{equation}
where $\Pi^{\text{ch}}(0,\vek{0})=-\frac{N_{\rm f}}{f_\pi^2}\Delta\phi(\Delta,\mu)$ which exactly cancels the tadpole contribution in Eq.~\eqref{eq:tad} with $\mu_8=0$, reflecting the Goldstone theorem. The coefficients $A$ and $B$ can be evaluated by employing the high-density approximation (at zero temperature), which consists in the replacement $\int\frac{d^3\vek{q}}{(2\pi)^3}\to N_0\int_{-\infty}^\infty d\ell$ with $\ell\equiv|\vek{q}|-\mu$ being the momentum measured with respect to the Fermi surface. This yields the result~\footnote{We note that to this order of approximation, the subtle problem associated with the momentum assignment to the two fermion propagators in the loop integral (see Ref.~\cite{Brauner:2008td}) is absent. This is because the one-dimensional $\ell$-integral is finite; any two momentum assignments differ by a shift which can be absorbed into a redefinition of the integration variable $\ell$.}
\begin{equation}
A=\frac{N_{\rm f}N_0}{2f_\pi^2},\qquad
B=\frac{N_{\rm f}N_0}{6f_\pi^2}.
\label{eq:coeff}
\end{equation}
In order for the NG boson Lagrangian to have the canonical form, we adjust the decay constant as $f_\pi^2=N_{\text{f}}N_0/2$. 

The knowledge of the diagonal elements of the polarization matrix is not sufficient to determine the dispersion relation or even to conclude that there are two massless modes. To that end, we need to evaluate the offdiagonal component for which we obtain the result at the leading order in the gradient expansion,
\begin{equation}
\Pi_{\text{mix}}(\omega,\vek{P})=-iC\omega,
\end{equation}
where $C=\Sqrt{3}n_8/(2f_\pi^2)$. In HDET, the color charge density takes the value $n_8=N_{\rm f}N_0\Delta^2/(2\Sqrt{3}\mu)$. This is suppressed by the small ratio $\Delta/\mu$ and thus belongs to the next-to-leading order in HDET. Nevertheless, we keep the term $C$ since the charge density $n_8$ is finite and can be evaluated without the high-density approximation. The polarization matrix in the $(\pi_4,\pi_5)$ sector then becomes, in the gradient expansion,
\begin{equation}
\begin{pmatrix}
-A\omega^2+B\vek{P}^2& -iC\omega\\
+iC\omega&-A\omega^2+B\vek{P}^2
\end{pmatrix}.
\label{casei}
\end{equation}
Upon diagonalization in the $(\pi_+,\pi_-)$ basis, this leads to $\Pi_\pm(\omega,\vek{P})=-A\omega^2+B\vek{P}^2\mp C\omega$, resulting in the dispersion relations
\begin{equation}
\omega_{\pi_+}=\frac{B}{C}\vek{P}^2,\qquad
\omega_{\pi_-}=\frac{C}{A}+\frac{B}{C}\vek{P}^2
\end{equation}
to second order in momentum. We can see that $\pi_+$, having the like charge as the medium, becomes a type-II NG boson with a quadratic dispersion relation, while $\pi_-$ with the opposite charge acquires a gap proportional to the color density $n_8$. This is in agreement with the general theorems concerning NG bosons in systems lacking Lorentz invariance~\cite{Nielsen:1975hm,Blaschke:2004cs,Brauner:2008td}. Also, it is clear that the system is stable at least at the quadratic order in the derivative expansion.


\subsubsection{Analysis for the case (ii)}

The calculation is basically the same as in the case (i), but we need to take into account the chemical potential differences, $\mu_r=\mu_g\ne\mu_b$. As before, we use the basis $(\pi_+,\pi_-)$, and thus need to analyze the polarization function $\Pi_+(\omega,\vek{P})=\Pi^{\text{ch}}_{\text{tad}}+\Pi^{\text{ch}}(\omega,\vek{P})-i\Pi_{\text{mix}}(\omega,\vek{P})$. It is easy to evaluate the long-wavelength limit of $\Pi^{\text{ch}}(\omega,\vek{P})[=\Pi^{\text{ch}}_{\Delta\Delta}(\omega,\vek{P})]$,
\begin{align}
\notag
\Pi_{\Delta\Delta}^{\text{ch}}(0,\vek{0})=&-\frac{N_{\rm f}}{f_\pi^2}\Delta\phi(\Delta,\mu_r)\\
\notag
&+\frac{\Sqrt{3}N_{\rm f}}{4f_\pi^2}\mu_8\Delta^2\int\frac{d^3\vek{q}}{(2\pi)^3}\Biggl[\frac{\tanh\left(\frac{\ell_b}{2T}\right)}{\ell^2+\Delta^2-\ell_b^2}\\
&-\frac{\ell_b\tanh\Bigl(\frac{\Sqrt{\ell^2+\Delta^2}}{2T}\Bigr)}{\Sqrt{\ell^2+\Delta^2}(\ell^2+\Delta^2-\ell_b^2)}\Biggr],
\label{eq:DD}
\end{align}
where $\ell=|\vek{q}|-\mu-\mu_8/(2\Sqrt{3})$ and $\ell_b=\ell+\Sqrt{3}\mu_8/2$ are the momenta for red/green quarks and for blue quarks measured from the Fermi surface. The first term in $\Pi_{\Delta\Delta}^{\text{ch}}(0,\vek{0})$ is cancelled by the tadpole contribution $\Pi^{\text{ch}}_{\text{tad}}$, but the second term survives. Therefore, in this case the curvature mass remains finite.

Let us take a slightly different approach to the problem. An explicit computation reveals that the function $\Pi_+(\omega,\vek{P})$ depends on $\omega$ only through the combination $\omega+\mu_b-\mu_r=\omega-\Sqrt{3}\mu_8/2$. This fact suggests that it would be more natural to perform the expansion about $\omega=\Sqrt{3}\mu_8/2$ for $\Pi_+(\omega,\vek{P})$ and similarly about $\omega=-\Sqrt{3}\mu_8/2$ for $\Pi_-(\omega,\vek{P})$. In fact, we can easily show that the offset of $\Pi^{\text{ch}}(\omega,\vek{P})\mp i\Pi_{\text{mix}}(\omega,\vek{P})$ at $\omega=\pm\Sqrt{3}\mu_8/2$ completely cancels the tadpole contribution so that
\begin{equation}
 \Pi_{+}(\omega=\Sqrt{3}\mu_8/2,\vek{0})=\Pi_{-}(\omega=-\Sqrt{3}\mu_8/2,\vek{0})=0.
\end{equation}
Therefore, we here perform the gradient expansion of $\Pi_{+}$ about $\omega=\Sqrt{3}\mu_8/2$. Up to second order in $\omega$ and $\vek{P}$ we find
\begin{equation}
\begin{split}
\Pi_{+}(\omega,\vek{P})=&A(\omega-\Sqrt{3}\mu_8/2)\\
&-B(\omega-\Sqrt{3}\mu_8/2)^2+C\vek{P}^2,
\end{split}
\end{equation}
where $A$ is proportional to the color density,
\begin{equation}
A=\frac{\Sqrt{3}}{4f_\pi^2}n_8(\Delta,\hat{\mu})=0.
\end{equation} 
This is nothing but the neutrality condition which determines the value of $\mu_8$. In the high-density approximation, this can be calculated explicitly, $\mu_8=-\Delta^2/(2\Sqrt3\mu)$. (This is actually a next-to-leading order result, for it is suppressed as compared to $\Delta$ by the factor $\Delta/\mu$.) One can also provide explicit integral formulas for $B$ and $C$; since they are rather complicated, we again use the high-density approximation with the result
\begin{equation}
\begin{split}
B&=\frac{N_{0r}N_{\rm f}}{2f_\pi^2}\frac{\Delta^2+2\mu_8^2/3}{\Delta^2}\simeq\frac{N_0N_{\rm f}}{2f_\pi^2},\\
C&=\frac{N_{0r}}{2f_\pi^2}\frac{N_{\rm f}}{3}\simeq\frac{N_0N_{\rm f}}{6f_\pi^2},
\end{split}
\end{equation}
where $N_{0r}\equiv{\mu_r^2}/{2\pi^2}$. Setting $f_\pi=\Sqrt{N_0N_{\rm f}/2}$ and the phase velocity $v_\pi=1/\Sqrt{3}$, the polarization functions acquire the final form
\begin{equation}
\Pi_\pm\simeq-(\omega\mp\Sqrt{3}\mu_8/2)^2+v_\pi^2\vek{P}^2,
\label{caseii}
\end{equation}
near $\omega=\pm\Sqrt{3}\mu_8/2$, $\vek{P}=\vek{0}$. 

We can see that under the color neutrality condition, the propagator of the charged NG modes acquires a double pole at $\omega=\pm\Sqrt{3}\mu_8/2$ which clearly indicates some kind of instability as already explained in Section~\ref{Subsec:fixedmu}. These shifted poles were recognized as the ``mass'' of the pseudo-Goldstone modes due to the``explicit symmetry breaking'' by $\mu_8$~\cite{Blaschke:2004cs,He:2005mp} since $\det{\Pi}(\omega,\vek{0})\propto(\omega^2-3\mu_8^2/4)^2$. Finally, we remark that, in accord with the discussion in Section~\ref{Subsec:instability}, the instability is also clearly seen in the curvature mass squared,
\begin{equation}
\frac{1}{V}\frac{\partial^2\Omega}{\partial\Delta_{1}^*\partial\Delta_{1}}=\Pi_\pm(0,\vek{0})=-\frac{3\mu_8^2}{4}<0.
\end{equation}


\subsubsection{Analysis for the case (iii)}

We finally consider the most general case with an arbitrary value of $\mu_8$ and the corresponding color density $n_8$, but with a proper account of the fluctuations in the color chemical potentials.

Let us start with the diagonal element of the polarization matrix, $\Pi_{\text{tad}}+\Pi^{\text{ch}}(\omega,\vek{P})$. In this case we need to take into account the contributions from $\Pi^{\text{ch}}_{\Delta\mu}$ and $\Pi^{\text{ch}}_{{\mu\mu}}$ in addition to $\Pi_{\Delta\Delta}^{\text{ch}}$ whose long-wavelength limit is given in Eq.~\eqref{eq:DD}. Computing all the contributions using the explicit expressions for quark propagators, we obtain, in the long-wavelength limit
\begin{multline}
\Pi_{\Delta\Delta}^{\text{ch}}(0,\vek{0})+\Pi_{\mu\Delta}^{\text{ch}}(0,\vek{0})+\Pi_{\mu\mu}^{\text{ch}}(0,\vek{0})\\ 
=-\frac{N_{\text{f}}}{f_\pi^2}\Delta\phi(\Delta,\mu_r)-\frac{3}{4f_\pi^2}\mu_8 n_8(\Delta,\hat{\mu}).
\end{multline}
Note that the above offset is completely cancelled by the tadpole contribution in Eq.~\eqref{eq:tad}. Thus, the diagonal entry of the polarization matrix can be expanded up to second order in $\omega$ and $\vek P$ as
\begin{equation}
\Pi_{\text{tad}}+\Pi^{\text{ch}}=-B\omega^2+C\vek{P}^2,
\end{equation}
where $B$ and $C$ are given by the same expression as in Eq.~\eqref{eq:coeff} up to the leading order in $\mu_8/\mu$. This is true regardless of the value of $\mu_8$  as long as the color chemical potentials fluctuate with the primary order parameter according to Eq.~\eqref{eq:parametrization}. In this case the curvature mass squared always vanishes,
\begin{equation}
\frac{1}{V}\frac{d^2\Omega}{d\Delta_1^*d\Delta_1}=\Pi_{\text{tad}}(0,\vek{0})+\Pi^{\text{ch}}(0,\vek{0})=0.
\end{equation}

In order to understand the low-energy behavior of the NG modes, we still need to evaluate the off-diagonal part of the polarization matrix, $\Pi_{\text{mix}}$; this consists of three parts $\Pi_{\text{mix}}^{\Delta\Delta}$, $\Pi_{\text{mix}}^{\Delta\mu}$, and $\Pi_{\text{mix}}^{\mu\mu}$. Each of the contributions takes quite a complicated form, but surprisingly putting them all together gives rise to the following simple formula at the lowest nontrivial order in the gradient expansion,
\begin{multline}
\Pi^{\Delta\Delta}_{\text{mix}}(\omega,\vek{P})+\Pi^{\mu\Delta}_{\text{mix}}(\omega,\vek{P})+\Pi^{\mu\mu}_{\text{mix}}(\omega,\vek{P})\\
=-i\frac{\Sqrt{3}}{2f_\pi^2}\omega n_8(\Delta,\hat{\mu})+{\mathcal O}(\omega^3,\omega P^2).
\end{multline}
This only depends on $\mu_8$ through the color density $n_8$. To summarize, the polarization matrix in the charged sector behaves at long wavelength as
\begin{equation}
\begin{pmatrix}
-B\omega^2+C\vek{P}^2 & -i\frac{\Sqrt{3}}{2f_\pi^2}n_8(\Delta,\hat{\mu})\omega\\
 i\frac{\Sqrt{3}}{2f_\pi^2}n_8(\Delta,\hat{\mu})\omega & -B\omega^2+C\vek{P}^2
\end{pmatrix}.
\label{caseiii}
\end{equation}
It is now obvious that only when $\mu_8$ is tuned such that the color density $n_8$ vanishes, the full set of five type-I NG bosons are recovered. 

\begin{figure}
\centering
\includegraphics[width=\columnwidth]{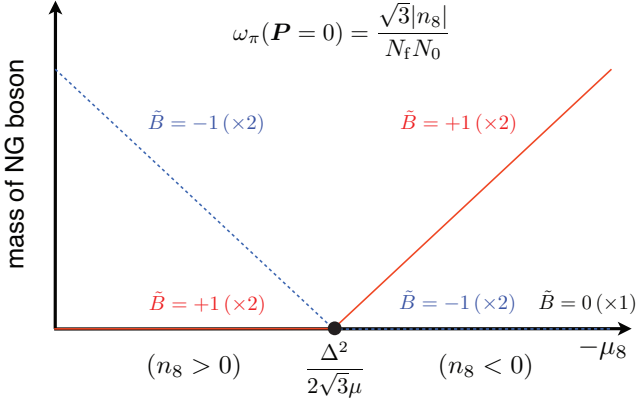}
\caption{(color online). The schematic picture of the NG boson masses as a function of the color chemical potential $\mu_8$. The neutral NG mode is always type-I and it is always gapless to the lowest order of approximation.}
\label{fig:ngmasses}
\end{figure}

In Fig.~\ref{fig:ngmasses} we show the schematic plot of the excitation gaps (masses) of the NG modes as a function of~$\mu_8$. There is one neutral NG boson which is  always type-I irrespective of the value of $\mu_8$ and the charge density of the system. The four charged modes with $\tilde{B}=\pm1$ are, on the contrary, sensitive to $\mu_8$. When $\mu_8>-\Delta^2/(2\Sqrt{3}\mu)$, the system is positively charged with color $n_8$ and the two type-II NG modes appear in the $\tilde B=+1$ sector; their antiparticles have the gap $\omega=\Sqrt{3}n_8/(N_{\rm f}N_0)$. On the other hand, if $\mu_8<-\Delta^2/(2\Sqrt{3}\mu)$, the system is negatively charged, and the quantum numbers of type-II NG modes and those of their massive partners are interchanged. At the neutrality point, $\mu_8=-\Delta^2/(2\Sqrt{3}\mu)$, the type-II NG modes change smoothly to the type-I NG modes.

\subsection{Effective Lagrangian for the charged NG modes}

The results for the dispersion relations of the charged NG modes can be conveniently encoded in an effective Lagrangian. Generally, the nonlinear effective Lagrangian can be constructed using the coset field $\mathcal U(\vek\pi)$ introduced in Eq.~\eqref{eq:calU}. A symmetry transformation $g\in\gr{SU(3)\times U(1)}$ acts on this field as $\mathcal U(\vek\pi)\xrightarrow{g}\mathcal U(\vek\pi')=g\,\mathcal U(\vek\pi)h^{-1}(\vek\pi,g)$, where $h(\vek\pi,g)$ is a suitable element of the unbroken subgroup $\gr{SU(2)\times U(1)}_{\tilde B}$.

A convenient way to ensure invariance of the effective Lagrangian is to introduce two matrix-valued fields
\begin{equation}
\mathcal T\equiv\mathcal UT_2\mathcal U^{T},\qquad 
\mathcal A\equiv\mathcal UT_8\he{\mathcal U}.
\end{equation}
These embody the existence of two order parameters, transforming in the antisymmetric rank-2 tensor representation and the adjoint representation, respectively. In the context of HDET, they were introduced in Eq.~\eqref{eq:parametrization}, see also Eq.~\eqref{adjoint} for the analogous construction within the toy model of Section~\ref{Sec:toymodel}. Under the action of $g$, these fields transform homogeneously as $\mathcal T\xrightarrow{g}g\mathcal Tg^T$ and $\mathcal A\xrightarrow{g}g\mathcal A\he g$.

As long as we are interested only in the dispersion relations of the charged NG modes, we can use any of the fields $\mathcal T,\mathcal A$ to construct the effective Lagrangian. (They lead to kinetic terms which are equivalent up to a redefinition of the parameters of the Lagrangian.) We choose $\mathcal A$ without loss of generality, and write down the most general $\gr{SU(3)\times U(1)}$-invariant Euclidean effective Lagrangian up to second order in the fields,
\begin{equation}
\mathcal L_{\text{eff},0}=\frac{4f_\pi^2}{3}\tr\left(\partial_\tau\mathcal A\partial_\tau\mathcal A+v_\pi^2\nabla\mathcal A\cdot\nabla\mathcal A\right),
\end{equation}
where the prefactor has been chosen just for convenience. Since there is an unbroken $\gr{U(1)}_{\tilde B}$ symmetry, we should allow for the possibility that it is endowed with a chemical potential. This is done by the replacement
\begin{equation}
\partial_\tau\mathcal A\to\partial_\tau\mathcal A-\mu_{\tilde{B}}[\tilde B,\mathcal A].
\end{equation}
Finally, the background $n_8$ charge density breaks the global color symmetry ``explicitly''. Since the charge density transforms in the adjoint representation, we have to add a term
\begin{equation}
\delta{\mathcal L}_{\text{eff}}=-\frac{8}{3}M_{\pi}^2f_{\pi}^2\tr\left(\mathcal AT_8\right),
\end{equation}
where the parameter $M_\pi^2$ measures the amount of explicit symmetry breaking, very much like the pion mass in the chiral perturbation theory of QCD. Expanding the action up to second order in $\vek{\pi}$ and discarding irrelevant terms, we arrive at the lowest-order action for the NG modes in the $(\pi_4,\pi_5)$ sector,
\begin{align}
\notag
\mathcal L_{\text{eff}}=&\tr\Bigl[\bigl(\partial_\tau\vek{\pi}-\mu_{\tilde{B}}[\tilde{B},\vek{\pi}]\bigr)^2+v_\pi^2(\nabla\vek{\pi})^2+M_\pi^2\vek{\pi^2}\Bigr]\\
\notag
=&[(\partial_\tau+\mu_{\tilde B})\pi_-][(\partial_\tau-\mu_{\tilde B})\pi_+]+v_\pi^2\nabla\pi_-\cdot\nabla\pi_+\\
&+M_\pi^2\pi_-\pi_+,
\label{eq:lag}
\end{align}
where now $\vek{\pi}\equiv \pi_4T_4+\pi_5 T_5$. The low-energy couplings $\mu_{\tilde{B}}$, $v_\pi^2$, and $M_\pi^2$ have to be computed in some microscopic model, as we have done above using HDET. Let us emphasize that $\mu_{\tilde B}$ is to be interpreted as an \emph{effective} chemical potential; it is not equal to the chemical potential of the blue quarks even though $\tilde B$ represents the operator of blue quark number up to an overall factor.

The Lagrangian~\eqref{eq:lag} implies the dispersion relations for the $\pi_\pm$ modes,
\begin{equation}
\omega_{\pi_\pm}=\sqrt{v_\pi^2\vek P^2+M_\pi^2}\mp\mu_{\tilde B}.
\end{equation}
Concretely, in the case (ii) where the fluctuations of color chemical potentials are ignored, we obtained $\mu_{\tilde{B}}=-\Sqrt{3}\mu_8/2$, $v_\pi^2=1/3$, and $M_\pi^2=0$, see Eq.~\eqref{caseii}. The instability of the $\pi_+$ mode, akin to the instability revealed in Section~\ref{Subsec:instability}, is now manifest.

The cases (i) and (iii) can be treated together; the expressions valid for the former can be obtained by setting $\mu_8=0$ in those for the latter. We obtained $\mu_{\tilde{B}}=\Sqrt3n_8(\Delta,\hat{\mu})/(4f_\pi^2)=\Sqrt3n_8(\Delta,\hat{\mu})/(2N_{\rm f }N_0)$, $v_\pi^2=1/3$, and $M_\pi=|\mu_{\tilde{B}}|$, see Eqs.~\eqref{casei} and \eqref{caseiii}. In the case (i) where color neutrality is not imposed, both $\mu_{\tilde B}$ and $M_\pi$ are nonzero and we find one type-II NG mode and one massive mode with the gap $2M_\pi$. On the other hand, in the case (iii) both parameters are zero and we find two type-I NG bosons with the phase velocity $v_\pi$.

On the general note, the above result can be interpreted as the background charge density in the fermion sector acting as a chemical potential for the modified baryon number. This provides some intuitive understanding of how the propagation of the NG modes is affected by the background charge. Putting together the collective modes in all (charged as well as neutral) sectors, the conclusion for the two physical cases (i) and (iii) is as follows. For $n_8\ne0$, there are three NG modes, two of which are type-II with $\tilde{B}=1$ and one type-I with $\tilde{B}=0$. When $n_8=0$, two gapped modes with $\tilde B=-1$, the antiparticles of the $\tilde B=1$ NG bosons, become gapless, and five type-I NG modes are correctly recovered.


\section{Conclusions}
\label{Sec:conclusions}

In this paper we analyzed in detail the issue of color neutrality in the 2SC phase of dense quark matter with particular attention to the spectrum of fluctuations of the order parameter. We showed explicitly that once color neutrality is imposed properly, the spurious instability as well as explicit breaking of the global color symmetry reported previously in literature disappear. To avoid confusion, we first addressed the problem in Section~\ref{Sec:NGmasses} using the same setting as in the previous publications. Only then, in Section~\ref{Sec:NGdispersions}, we provided a model-independent approach to the dispersion relations of the NG modes. While we used the 2SC phase as a specific example, it is obvious that our conclusions are valid more generally. Apart from other color-superconducting phases, they apply equally well to all systems with a set of conserved charges, some of them being demanded to be zero in the equilibrium.

It should be pointed out that the problem investigated in this paper is to some extent artificial. In the full QCD, color neutrality is automatically guaranteed by the gauge dynamics.  Moreover, the collective modes discussed here are all colored and thus do not correspond to physical states in the spectrum; they will eventually be absorbed by the gluons by means of the Higgs--Anderson mechanism. It was shown that the collective excitations, the NG modes, play an important role in establishing gauge invariance of the Meissner effect~\cite{Nambu:1960tm}. We thus believe that our work provides a first step towards the proof of gauge invariance of the color Meissner effect in color superconductors, which has not been demonstrated satisfactorily so far. We leave a deeper consideration of this issue to future work.


\begin{acknowledgments}
Our current understanding of the problem, as presented in this paper, has been shaped by discussions with numerous colleagues. We would like to express our gratitude to those who influenced it most strongly, namely J.~O.~Andersen, M.~Buballa, X.-g.~Huang, K.~Fukushima, D.~H.~Rischke, and especially I.~A.~Shovkovy. The research of T.B.~was supported by the Sofja Kovalevskaja program of the Alexander von Humboldt Foundation.
\end{acknowledgments}


\bibliography{references}

\end{document}